# Perspective on Coupled Colloidal Quantum Dot Molecules

Somnath Koley[1,2,†], Jiabin Cui[1,2,†], Yossef E. Panfil[1,2,†], Uri Banin[1,2]*

1 Institute of Chemistry, The Hebrew University of Jerusalem, Jerusalem 91904, Israel.

2 The Center for Nanoscience and Nanotechnology, The Hebrew University of Jerusalem, Jerusalem 91904, Israel.

† These authors contributed equally to this work.

* Corresponding author. Email: uri.banin@mail.huji.ac.il (U.B.)

**Abstract:** Electronic coupling and hence hybridization of atoms serves as the basis for the rich properties for the endless library of naturally occurring molecules. Colloidal quantum dots (CQDs) manifesting quantum strong confinement, possess atomic like characteristics with *s* and *p* electronic levels, which popularized the notion of CQDs as artificial atoms. Continuing this analogy, when two atoms are close enough to form a molecule so that their orbitals start overlapping, the orbitals energies start to split into bonding and anti-bonding states made out of hybridized orbitals. The same concept is also applicable for two fused core-shell nanocrystals in close proximity. Their band edge states, which dictate the emitted photon energy, start to hybridize changing their electronic and optical properties[1-2]. Thus, an exciting direction of "artificial molecules" emerges leading to a multitude of possibilities for creating a library of new hybrid nanostructures with novel optoelectronic properties with relevance towards diverse applications including quantum technologies.

The controlled separation and the barrier height between two adjacent quantum dots are key variable for dictating the magnitude of the coupling energy of the confined wavefunctions. In the past, Coupled double quantum dot architectures prepared by molecular beam epitaxy revealed a coupling energy of few meV, which limits the applications to mostly cryogenic operation. Realization of artificial quantum molecules with sufficient coupling energy detectable at room temperature, call for the use of colloidal semiconductor nanocrystal building blocks. Moreover, the tunable surface chemistry widely opens the predesigned attachment strategies as well as the solution processing ability of the prepared artificial molecules, making the colloidal nanocrystals as an ideal candidate for this purpose. Despite several approaches demonstrated enabling the coupled structures[1,3], a general and reproducible method applicable to a broad range of colloidal quantum materials is demanded for systematic tailoring of the coupling strength based on a dictated barrier

This report addresses the development of *nanocrystal chemistry* to create coupled colloidal quantum dot molecules and to study the controlled electronic coupling and their emergent properties. The simplest nanocrystal molecule, a homodimer formed from two core/shell nanocrystal monomers, in analogy to homonuclear diatomic molecules, serves as a model system. The shell material of the two CQDs is structurally fused resulting in a continuous crystal. This lowers the potential energy barrier, enabling the hybridization of the electronic wavefunctions. The direct manifestation of the hybridization reflects on the band edge transition shifting towards lower energy and is clearly resolved at room temperature. The hybridization energy within the single homodimer molecule is strongly correlated with the extent of structural continuity, the delocalization of the exciton wavefunction and the barrier thickness as calculated numerically. The hybridization impacts the emitted photon statistics manifesting faster radiative decay rate, photon bunching effect and modified Auger recombination pathway compared to the monomer artificial atoms. Future perspectives for the nanocrystals chemistry paradigm are also highlighted.

## 1. Introduction

Colloidal Quantum Dots (CQDs) are semiconductor nanocrystals containing hundreds to thousands of atoms arranged in a crystal structure of the bulk semiconductor.[4,5] When the crystal is smaller than the exciton Bohr radius of the semiconductor, the energy levels, especially near the band edge, blue shift and become discrete because of the quantum confinement effect.[6–8] This allows to tune the electron and hole states as a function of size and shape. Core/shell quantum dots are particularly interesting as the shell material passivates the dangling bonds on the surface and the shell acts as an electronic potential barrier for the charge carriers keeping them away from the traps on the surface.[9] Extensive studies over last three decades focusing on the structure, surface and interface achieved superior optical and electronic properties leading to applications of quantum dots in displays, photodetectors, solar harvesting and as bio-markers to name a few.

How two adjacent CQDs interact in close proximity and the ways to couple the neighboring particles are important questions in realizing nanocrystal based devices. Specifically utilizing the artificial atom character of the quantum dots (to be discussed later), hybridization of the confined wavefunctions in two dots is anticipated.[10] In the following, we present the overview of the latest findings and current progress on CQD molecules. First, a brief introduction of the synthetic strategies, followed by coupling attributes via wavefunction hybridization in the coupled homodimer CQD molecules, will be discussed. Next, the photophysical aspects of the coupled CQDs, down to single particle level, are presented. Finally, we provide a perspective on the challenges and future opportunities in this field towards potential applications.

## 2. From artificial atom to artificial molecules

Due to the quantum confinement effect, the CQDs present a transition state between the solid-state and molecular regimes with discrete energy levels and electronic states, and are often described as 'artificial atoms'.[4] This analogy was clearly revealed experimentally, where the spherical shape CQDs present atomic-like symmetries and degeneracies for the electronic states and wavefunctions, for example '*s*' and '*p*' states.[11,12] As depicted in Figure.1a-g, the different extent of the conduction band (CB) *s* and *p* states of InAs/ZnSe core/shell nanocrystals were directly probed by scanning tunneling microscopy (STM). Bias-dependent current imaging delineated the atomic-like states of the CQDs by measuring the $dI/dV - V$ spectrum (Figure.1a). Current images at different bias values directly revealed the nature of the *s* state at lower bias and *p* states at higher bias (Figure.1b-g).

The observations of the artificial atom character of CQDs, inspired the formation of quantum dot molecules where the creation of hybridized excitons can impose novel synergistic functionalities, relevant to numerous applications including in quantum technologies. This requires the controlled placement of a second quantum system in close proximity.[13] The chemical bond serves as a basis of molecule formation with electronically coupled atomic states, enabling tunneling of the electrons and delocalization throughout the new hybridized molecular state.[2] Similarly, close proximity of two quantum dots with a minimal barrier engenders hybridized electron wavefunction delocalized throughout the double quantum dot structure, forming a quantum molecule. In the following sections we will describe the advancement, hurdles and development of a general strategy for coupling two quantum dots into an artificial molecule. (Figure 1h).

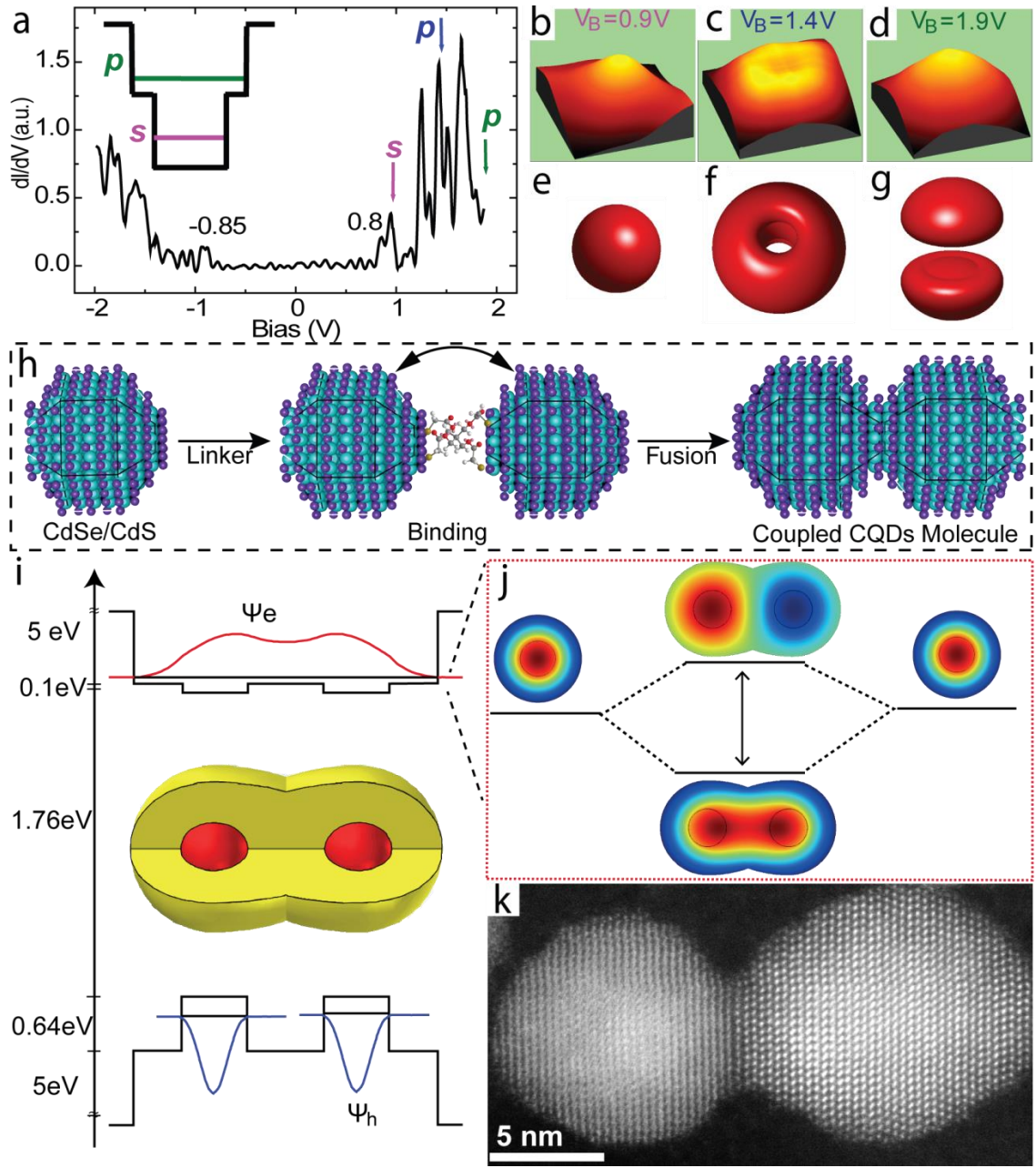

**Figure 1.** (a) Scanning tunneling spectrum acquired for the wavefunction imaging of InAs/ZnSe core/shell QDs. (b-d) Current images obtained at different bias values denoted by arrows in (a). (e-g) Isoprobability surfaces, showing $s^2$ (e), $p_x^2 + p_y^2$ (f), and $p_z^2$ (g). Adapted from ref [14], Copyright 2001 American Physical Society. (h) Simplified scheme for the synthesis of coupled artificial molecules formed from two core/shell CQDs monomers. Following the formation of dimers connected via organic linker, thermal fusion creates the coupled CQD molecule. (i) The potential energy landscape and a cross-section of the calculated first electron and hole wavefunctions in CdSe/CdS CQD molecules. (j) Calculated bonding and antibonding 2-

dimensional electron wave-functions without Coulombic interaction. (k) HAADF-STEM image for the CQDs homodimer molecule. Adapted from ref 1, Copyright 2019 Springer Nature.

## 3. Strategies to fabricate Artificial Quantum Molecules

The first attempts to hybridize the exciton wavefunction were made with epitaxially grown and self-assembled quantum dots. The close proximity of two quantum dots led to hybridized wavefunctions, that respond to perturbation from an external electric field, as studied at single-particle level.[15,16] In MBE grown QD molecules, along with the lack of solution processability, cryogenic operation remains a limitation owing to the typically small hybridization energies.[11,17,18]

For colloidal quantum dots, up to now, the successive ion layer adsorption and reaction (SILAR)-based method was used to form coupled structures such as core-multishells.[19,20] Double-QDs at nanorod apexes were also reported.[21–23] With controlled precursor reactivity and injection rate, growth of additional QDs could be tailored on the apex of dot-in-rod structures for example, and yielded a coupled system.[21] In these structures, dual emitting nanocrystals were demonstrated via spatial separation of the carrier wavefunction, with simultaneous interaction of the exciton via dipole or tunnel coupling. Wavefunction mixing in rod-couples lowering the emission polarization has also been demonstrated.[3] Due to the specific materials architectures and types combined with lattice mismatch and surface strain, the precise control of coupling parameters are limited by these strategies.

Utilizing the surface chemistry of the CQDs, dimer and oligomeric structures were prepared using nanoscale templates, including organic polymers,[24] biomolecules,[25] and nanotemplates (polystyrene and $SiO_2$ particles).[26] The linker/templates were utilized to shorten the distance between the building blocks to facilitate inter-particle interaction. On top of this, the template-free methods, utilizing dipole-dipole interaction,[27,28] ligand exchange,[29] solvent evaporation,[30] and electrostatic interaction,[26] were also used to assemble CQDs in close proximity. The ensemble of closely packed CQDs forming superlattice or molecular-like aggregates manifested the inter-dot interaction and coherent exciton dynamics at room temperature.[31,32] Even with a short cross linker that minimizes the distance between two CQDs, significant energy barrier is still imposed by the interfacial organic ligands and linkers, restricting the wavefunction hybridization. However, a dipolar non-radiative resonance energy transfer interaction can still be efficient with timescale down to tens of picoseconds.[30,31,33] These reports emphasized the distinct nature of the exciton kinetics compared to isolated quantum system in ambient conditions, driving further the motivation to generate and study isolated colloidal quantum dot molecules.

Fusion of nanocrystals, via oriented attachment (OA) provides a solution to this interfacial organic ligand barrier. During OA, a single nanocrystal is formed from two adjacent nanoparticles based on the formation of the atomically matched bond between the two specific facets, governed by the minimization of the highest surface energy of the most reactive facets.[34–37] The controlled OA is widely utilized for the formation of nanocrystals with various novel architectures, including nano-wire nanochains, helical, branched and nanorings.[24] Various methods have been developed to fuse the nanocrystals *via.* thermal annealing,[38] pressure compression,[39] and pulsed laser annealing.[40] In a vision to hybridize the wavefunctions, this method was applied recently to Pb-chalcogenide nanocrystals. Notably, Beard and coworkers reported the broadening of the 1S exciton peak owing to wavefunction hybridization in PbSe fused dimers with a splitting of the degenerate energy levels of up to 140 meV.[41] Alivisatos and coworkers have analyzed the possible crystallographic defects at the interface of PbTe[42] and CdSe [43] nanocrystal arrays due to misfit dislocation, twinning or grain boundaries, and stacking forces. The attachment of the CQD facets forming a continuous lattice eliminates the ligand barrier facilitating the carrier wavefunction overlap while a rigorous control of the interfacial defects should be addressed. However, the oriented attachment of the nanocrystal facets in free solution often yields uncontrolled higher order oligomeric structures limiting the formation of isolated controlled dimers.

A template based approach overcame these limitations, where the selective dimer preparation is performed on a template thus reducing the chance of higher order oligomerization with high yield of the desired dimer structures.[1] As demonstrated in the following sections, by choosing the core/shell CQDs as artificial atoms where the continuous shell lattice embraces the core CQDs, eliminates the interfacial ligand barrier while the shell thickness itself defines the barrier height. This general approach offers the ability to prepare the fused nanocrystal molecules with wide variety of size, structure and composition, predesigned and tested computationally. As depicted in Figure 1i, the quantum-mechanical calculation for a CdSe/CdS homodimer molecule, reveals the hybridization of the $1S_e$ electron wavefunctions forming a bonding and anti-bonding states with corresponding wavefunctions presented in the Figure 1j, given by in-phase and anti-phase superpositions of the monomer wavefunctions, respectively (Figure 1k).[2]
6

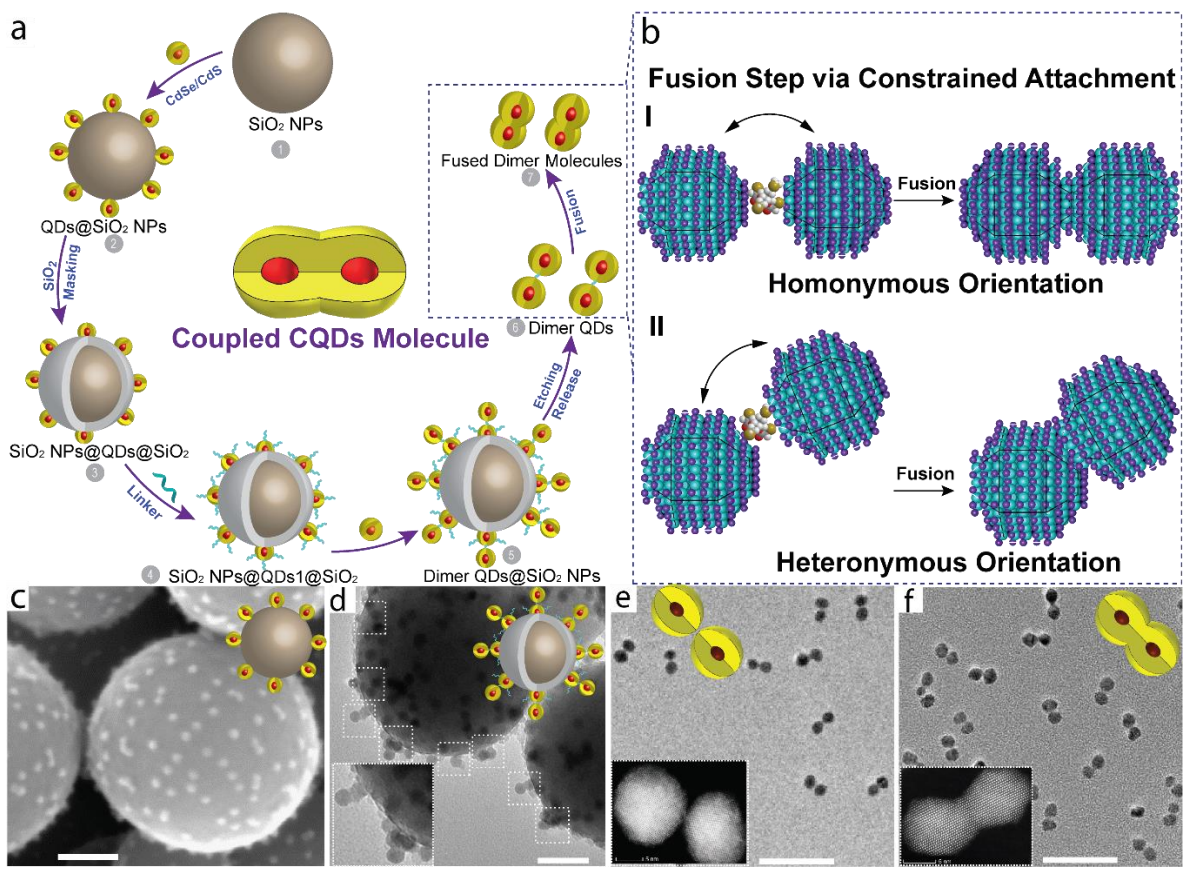

**Figure 2.** (a) Scheme for the formation of the coupled CQDs molecules utilizing SiO$_2$ particles as template. (b) The structural model for constrained attachment in CdSe/CdS CQD molecules with homonymous and heteronymous plane orientations during fusion . The TEM images of the (c) QDs@ SiO$_2$, (d) dimer QDs@ SiO$_2$, (e) non-fused dimer, and (f) the fused dimer. The insets in (e-f) are the related HAADF-STEM images. Adapted from ref 1, Copyright 2019 Springer Nature.

Sub-micron sized nanoshpheres have been utilized as the template as they offer high surface area and flexibility to tune the surface chemistry for anchoring small nanocrystals and manipulating their surface chemical functionality while anchored on the template itself.[44,45] The first step for the preparation of the CQD molecules is to prepare a cross linked CQD dimer on the template (Figure 2a). Surface functionalized silica particles first accommodate the core/shell CQDs (Figure 2c), following the controlled coverage of the unreacted silica anchor sites and a portion of the anchored CQDs. This masking step simultaneously reduces the possibility of both generating unreacted monomers and higher order oligomeric species along with the desired dimer. The second CQD can easily be anchored to the first CQD with a cross linker precisely, to form a cross-linked homodimer on the template surface (Figure 2d). The dimers are then detached from the template by etching the silica spheres (Figure 2e). These cross-linked homodimers however did not shift the 1S exciton peak lacking the wavefunction hybridization due to the high energy barrier imposed by the organic ligands and cross-linker at the interface.

A thermal induced fusion reaction is then applied to attach the facets of the wurtzite nanocrystals forming a continuous shell enabling the wavefunction tunneling throughout the dimer leading to a red-shift and broadening of the original CQDs first exciton peak (Figure 2f). Optimization of the fusion conditions, including the fusion time, temperature, and excess ligands, play a significant role in determining the dimer quality with a tradeoff between efficient fusion and ripening of the dimer structure. In the etching process to release the CQDs from silica, the removal of surface ligands of the CQDs may lead to the aggregation of the nanoparticles and thus, the addition of small amounts of oleylamine and oleic acid is utilized to stabilize the dimer structure and avoid the formation of the trimer, chains and aggregated structures during fusion. Importantly, the fusion temperature of 180-240 °C enables the efficient annealing of the continuous nanocrystal and also removes part of the surface and interfacial defects (Figure 2e-f). This approach holds the generality as successfully applied on the CQD nanocrystals with a variety of size and shapes with prior knowledge of the crystal structure, phase and surface chemistry. The dimer properties governing the extent of coupling in an artificial CQD molecule can be directly tailored with choice of monomer "artificial atoms" and fusion conditions (Figure 3). The tuning of the barrier via either altering the shell-thickness with sub-nm precision, or changing the neck width by tuning the fusion reaction conditions will influence the exciton tunneling efficiency. All these variations are approachable by the discussed synthetic method. Furthermore, the utilization of the well-developed wavefunction engineering in core/shell CQDs allows us to design the selective delocalization of electron and/or hole wavefunctions in the fused CQD molecules. While, in type-I core/shell CQDs, the charge carrier wavefunctions are tightly confined in the core; in quasi type-II CQDs, one of the carriers delocalizes to the shell, which systematically can control the extent of hybridization in fused dimers. An obvious presence of unreacted and aggregated species can be filtered out by well-developed separation techniques to enhance batch uniformity. Hence, from the synthetic point of view, this methodology appears to be simple and efficient to yield selectively fused CQD molecules with diverse size, shape and composition.

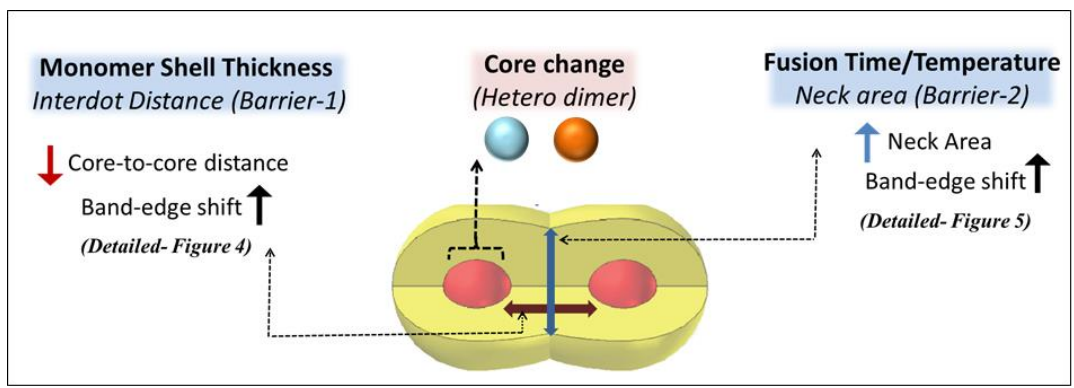

Figure 3. Possible alteration of band-edge properties affordable via simple synthetic modifications in fused CQD molecules.

## 4. Structural characterization of fused CQD homodimer molecule

During the masking step (step 3, scheme 2a), the CQDs become immobilized and only the top hemisphere sites remain available for the further chemical functionalization. While strong tetrathiol cross-linker molecules and successively the second CQDs are anchored to create linked dimers, the attachment is dictated by the facets bound with the cross-linker, which limits rearrangement of the CQDs in the fusion step unlike the typical oriented attachment. Thus, a constrained attachment with limited rotational freedom, fuses the facets quite strongly but generates dimer structures with both homonymous and heteronymous plane orientations (Figure.2b). High angle annular dark field (HAADF) scanning transmission electron microscopy (STEM) provides the precise information about the important parameters governing electronic coupling such as core location, center-to-center distance, and interface integration in the fused CQDs molecules.

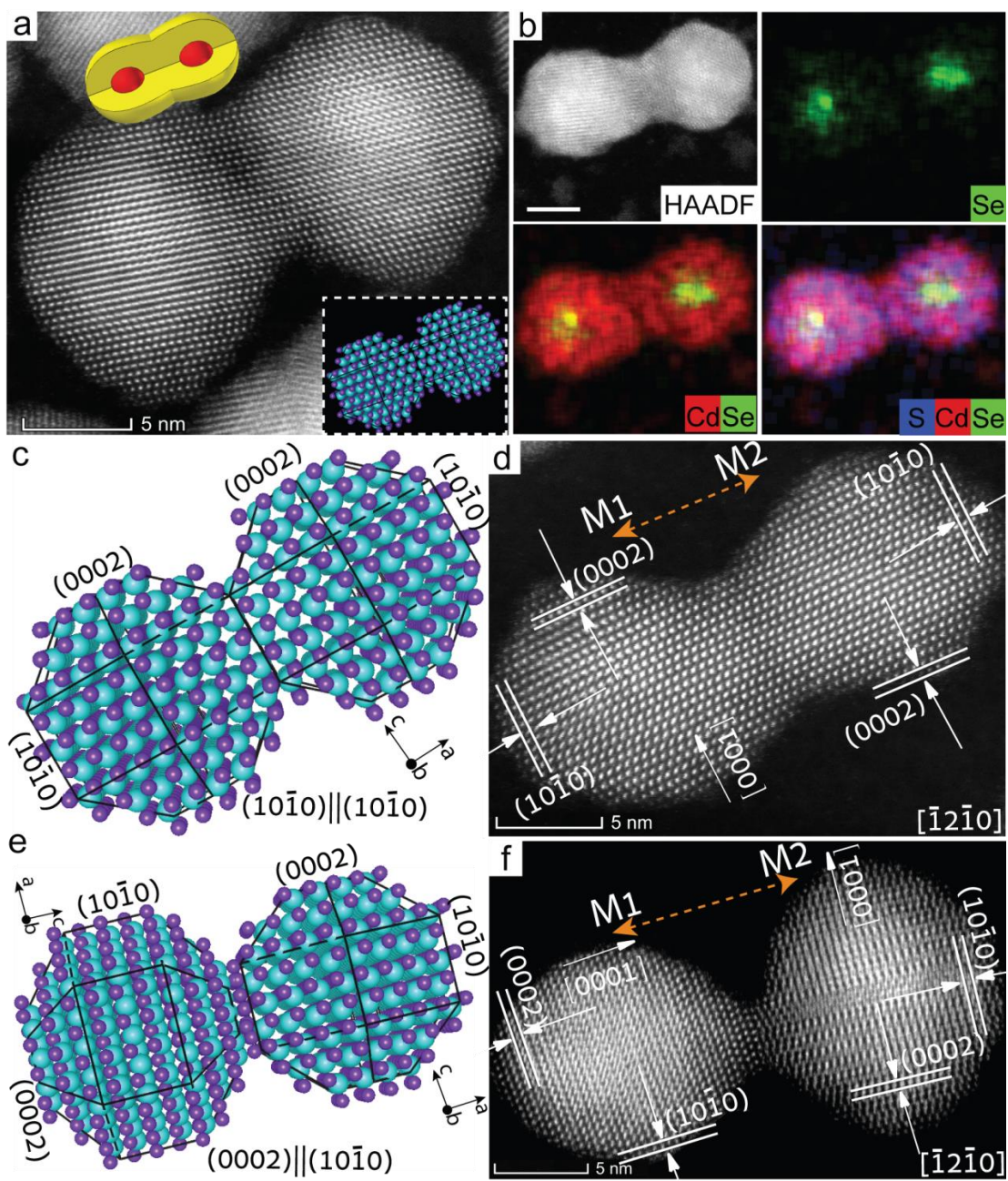

**Figure 4. Structural and chemical mapping of CQD molecules:** The HAADF-STEM image (a) and HAADF-EDS elemental mapping (b) of the coupled CdSe/CdS molecules. The atomic structure model and HAADF-STEM image respectively for (c-d) homo-plane attachment and (e-f)) hetero-plane attachment in fused dimers. For homo-plane attachment, the identical (10$\bar{1}$0) faces of M1 and M2 remain parallel, while for the hetero-plane attachment the parallel crystallographic faces are different [(0002) and (10$\bar{1}$0)], as observed under the same zone axis of [$\bar{1}$2$\bar{1}$0]. Adapted from ref 1, Copyright 2019 Springer Nature.

HAADF-STEM measurements (Figure 4), reveal a continuous lattice between two adjacent CQDs upon fusion forming the structurally coupled CQDs molecules (Figure 4a). HAADF-EDS elemental mapping clearly shows the fusion of the shell while keeping the cores intact (Figure 4b). Continuous distribution of cadmium (both in core and shell) and sulfur (only in shell) are identified along the dimer axis. Along the same line, selective regions of the selenium (only in core) are clearly identified signifying the cores locations.

The constrained attachment results in the homo-plane and hetero-plane fusion relationships as observed by the HAADF-STEM (Figure 4c-f), well-correlated with the simulated atomic structure model. As the specific facets of the wurtzite CdSe/CdS QDs possess (0002), (10$\bar{1}$0), and (10$\bar{1}$1) planes, several attachment scenarios emerge during the formation of the coupled CQDs molecules, unlike the typical OA process. In homo-plane attachment cases, both CQD monomers of a fused pair are projected under the same zone axis without any dislocation along the interface, leading to a high fusion area. This facilitate stronger tunnel coupling strength than in a dimer with low fusion area expected in the hetero-plane attachment cases.

## 5. Hybridization and electronic coupling effects in homodimer CQD molecules

While the attachment of two monomers to form a cross-linked dimer does not change band-edge transitions, the fusion step alters significantly both the absorption and emission features (Figure 5a).[2] Upon fusion, the only barrier which remains is the CdS shell, separating the two cores. However, when the barrier is thick and the cores are larger like in the case of 1.9nm/4nm core-radius/shell-thickness, the two electron wavefunctions on each of the cores remain localized in the big cores and along with the thick barrier in between the cores, this leads to a small red-shift of only 1-2 meV (figure 5b). As the cores become smaller, the electron wavefunctions start to delocalize the CdS shell and along with the thinner barrier thickness, the electronic states on each of the cores start to hybridize. This is well manifested in the red shift of the 1.4nm/2.1nm and the 1.2nm/2.1nm core-radius/shell –thickness, by 11 meV and 13 meV, respectively (figure 5b).[1]

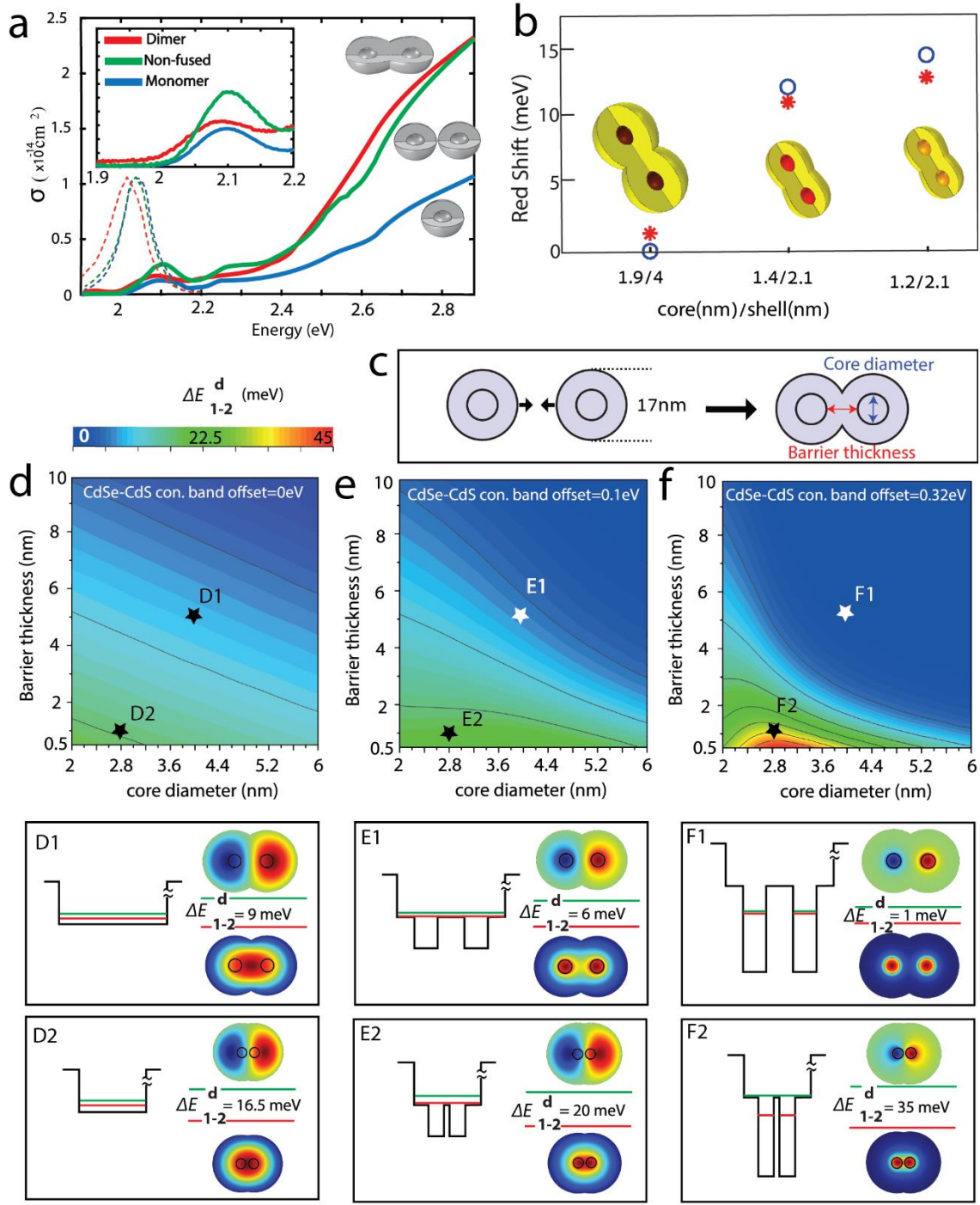

**Figure 5.** (a) Absorption cross section (solid-lines) and normalized emission spectra (dashed-lines) for the monomer (blue), nonfused dimer (green), and fused dimer (red). Adapted from ref 2. (b) Calculated (red-asterisk) and experimental (blue-circles) band-edge red-shift of monomer-to-respective-homodimer structures for different CQD molecules. Adapted from ref 1, Copyright 2019 Springer Nature. (c) The barrier width is controlled by overlapping the two outer spheres of the core/shell CQDs. (d) Contour plot

of the energy splitting between the symmetric and antisymmetric states as a function of barrier thickness and core diameter in: (d) 0 eV, (e) 0.1 eV and, (f) 0.32 eV CB offsets. The eigen-energies of the symmetric state (red-line) and antisymmetric state (green-line) for the electron with respect to the potential energy landscape along with corresponding wave-functions are presented for the points D1-F2 in the contours. Adapted from ref 2. Copyright 2019 American Institute of Physics.

To test the influence of the core diameter and the barrier thickness and to find general design rules for the dimers, we have simulated the dimer structure using Comsol Multi-Physics and solved the electron and hole Schrödinger equations using the effective-mass approximation. The conduction (CB) and valence band (VB) offset between CdSe to CdS, which serve in our system as the barrier height, is under debate in the literature and ranging between 0 eV-0.3eV for the CB offset, and we show results for 3 representative values: 0 eV, 0.1 eV and 0.32 eV. In any of the above mentioned cases, the VB offset is quite large and together with the heavier hole effective-mass, no significant hybridization is expected for the hole. So from now on we will concentrate on the electron energy levels.

The effect of the barrier thickness is highlighted by taking fixed monomers with overall dimer diameter of 17nm varying the overlap between them, in addition to tuning core sizes. As expected, for thicker shell barrier and larger core diameter, the hybridization energy $E_{1-2}^d$, which is the energy difference between the bonding and anti-bonding states, decreases. In addition, the decrease in the hybridization energy $E_{1-2}^d$ is more pronounced in case of larger band-offset because of the higher barrier height (points D1, E1 and F1 in figure 5d-f). Interestingly, in the case of a thin barrier and small core diameters the trend is opposite (points D2, E2 and F2 in figure 5d-f)[2].

Another crucial parameter which can greatly influence the hybridization energy is the neck formed between the monomers upon fusion (Fig 6a). The overlap diameter was varied between 0 nm, as in non-fused dimer, to 7 nm which resembles a totally fused dimer. A general trend of increasing hybridization energy with filling the neck prevails considering the CB-offsets values of 0.1eV and 0.32eV. This confirms that for small neck that arises with inefficient fusion, the neck constriction between the monomers acts as a potential barrier, suppressing the hybridization (Figure 6b-c). However, in the case of 0.32eV CB-offset the change in the hybridization energy is moderate, changing up to 6 meV with complete neck filling, compared to more pronounced change up to ~20meV for a 0.1eV CB-offset (Fig 6a).

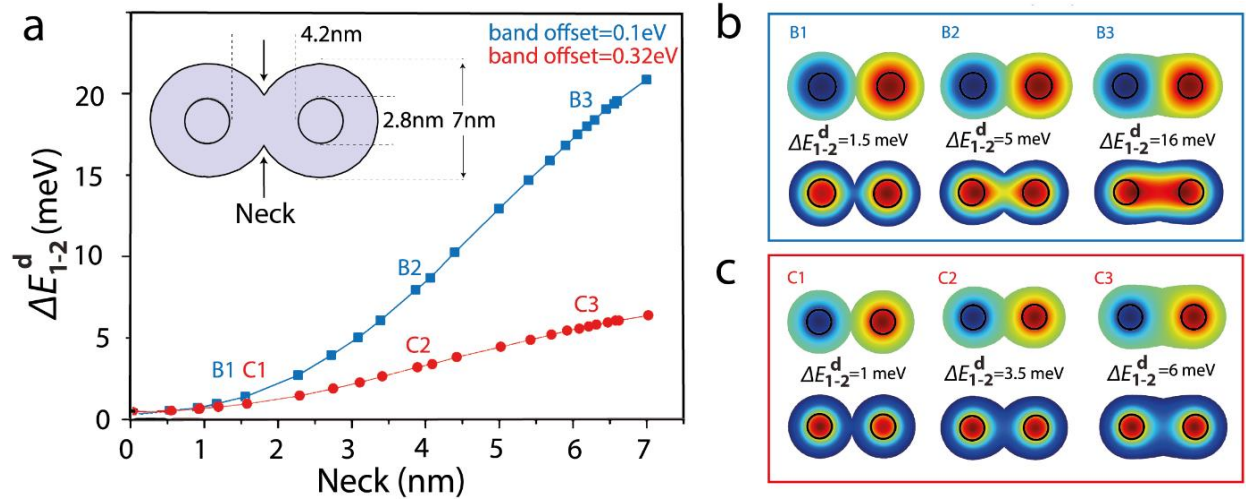

**Figure 6.** (a) Energy difference between the symmetric and antisymmetric states as a function of the neck diameter for 0.1 eV (blue), and 0.32 eV (red) CB offset in 1.4/2.1 homodimer. (b-c)The wave-functions of the symmetric (bottom) and antisymmetric (top) electronic states and the energy difference between them in three points: which refer to 1.5 nm, 4 nm, and 6.2 nm neck thickness, respectively, for (b) 0.1 eV band offset (B1–B3) and (c) for 0.32 eV band offset (C1–C3). Adapted from ref 2. Copyright 2019 American Institute of Physics.

The difference can be realized from the wavefunctions of the bonding and anti-bonding states for the two CB offsets. In case of 0.1 eV CB offset the electron wavefunction is delocalized in the CdS shell and hence the neck greatly influences the hybridization (B1, B2, and B3 in Figure 6b). However, for 0.32 eV CB offset the electron wavefunction remains localized in the core region and the neck restriction has less influence on the hybridization energy (C1, C2, and C3 in Figure 6c). It should be noted that the hybridization energy is not the only factor which determines the magnitude of monomer-to-dimer red-shift. The electron-hole Coulomb interaction greatly influences the red-shift and we should expect a moderate magnitude considering the Coulomb term.[2]

## 6. Coupling mechanisms in homodimer CQD Molecules

In a homodimer CQD molecule various coupling mechanisms, on top of the aforementioned hybridization, may become relevant. Resonant dipolar coupling or charge transfer interactions are possible, resulting in novel redistribution of the charge carriers in CQDs molecules. Single particle emission studies prove to be a powerful means to address these different possibilities. Individual monomer CQDs undergo the widely studied on-to-off blinking under photoexcitation as a result of exciton pair recombination and three carrier non radiative Auger recombination, respectively (Figure 7a).[46] This phenomena is typically indicated by a long mono-exponential radiative decay from the former bright state and a much faster decay from the latter dark state (Figure 7c). At the single exciton regime, the fused CQD molecule rather exhibit a flickering of emission intensity without a clear on or off state (Figure 7d), associated with a faster multi-exponential decay (Figure 7f). The quenching of the radiative decay time is observed and may be associated both with charge transfer and fast energy transfer processes between two CQDs, which occur before the successful recombination of the electron and hole within a single CQD. The dipolar fluorescence resonant energy transfer (FRET) remains a contributing factor for the faster decay rate as the organically linked non-fused dimer, where a tunneling of electron through the high barrier is less prominent, also exhibits some quenching (to smaller extent than the fused dimer). The non-radiative off state, obvious in monomer with the characteristic short lifetime designated to dark trions, was not observed in the studied dimers, as the extracted lifetime from the lowest intensity tags from single dimer remains significantly longer than monomers. This indicates the presence of a competitive pathway (*viz.* tunneling) which restricts fast Auger recombination to occur within one of the dots. Subsequently, owing to larger volume, the dimer can accommodate and stabilize trions efficiently by reducing the Auger rate itself enhancing the lifetime of the charged configuration (Figure 7 d, f).

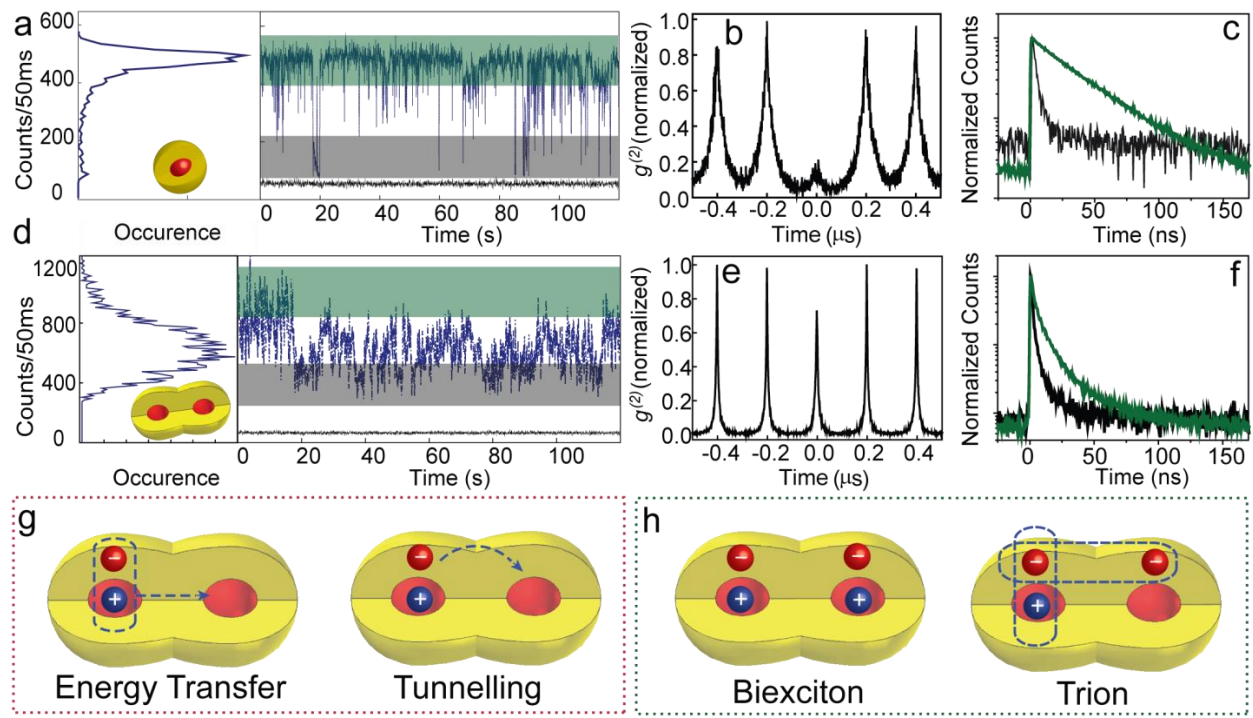

**Figure 7.** Photo-physical aspects of CQD molecules. Fluorescence time traces (a&d), second order photon correlation (b&e), Luminescence state dependent lifetimes (c&f) respectively from a single monomer (upper panel) CQD and homodimer CQD molecule (lower panel). (g) Quenching mechanism via inter-particle interaction and (h) novel multicarrier configurations in CQD dimer molecule. Adapted from ref 1, Copyright 2019 Springer Nature.

Thus, novel and interesting multicarrier configurations are possible in fused CQD molecules. The non-emissive off or weakly emissive gray states in colloidal CdSe/CdS CQDs arises due to efficient Auger recombination of the photo-generated trion states in thin shell CQDs. Significant reduction in Auger process was achieved in giant shell CQDs, where the spatial separation between the charge carriers reduces the overlap, and slows down the Auger rate.[47,48] The CQD molecule can offer new types of multicarrier configurations where the extra electron can tunnel to the second core and thus increase the trion lifetime due to spatial separation as compared to monomers. A bi-exciton configuration accommodating each exciton pair in either cores is an additional interesting feature that can be achieved in CQD molecules. A significant increase in biexciton quantum yield in CQD molecule revealed by the photon correlation statistics is shown in Figure 7e. At low intensity pulsed excitation, a near perfect photon antibunching at the zero time delay is observed in monomers, as the biexciton quantum yield in a thin shell CQD is lower owing to fast Auger decay.[49] On the other hand, a reduction in Auger rate enables significant photon bunching in the CQD molecule with a biexciton to exciton quantum yield ratio of up to 0.6 (Figure 7e) compared to monomer CQD revealing the value <0.1 (Figure 7b).[50] The realization of the new biexciton configuration is of importance towards utilization of CQD molecules in lasing applications.

## 7. Conclusion and perspectives

Pre-designed and on-demand electronic coupling strength is the key advantage with the coupled artificial molecules formed from core/shell building blocks by wet chemical approaches, that provide numerous opportunities for rational variation of the confined wavefunctions, along with controlling the extent of hybridization. Tuning the core and shell parameters i.e. the size, composition and thickness, can feasibly create a wide variety of heterostructured artificial molecules, with potential towards diverse emerging applications (Figure 8). (a) In a fused dimer molecule, the core size variation can manifest dual color emission from a single quantum system. The ability to tune and switch the emission color conforming to display applications are achievable in the CQD molecules by tailoring the core size and composition without changing the synthetic procedure.

(b) In type-II core/shell CQDs the staggered band alignment allows extensive delocalization of the one type of carrier to the shell. Higher degree of delocalization can facilitate the transfer of charge carriers to the neighboring CQDs. With a dictated barrier, a systematic alteration of rapid charge carrier transfer is relevant to the quantum information application where the coupled CQDs may serve as Qbits. This relies on the superposition of the charge carrier states, coulomb-interaction and coherent tunneling, those remain interesting topics of study for the near future in single dimer, a system which exhibit coupling energy resolvable at room temperature (>10meV)

(c) Through band-structure engineering, a long-lived charge carrier that is of relevance for light harvesting applications, can rationally be designed and achieved in a CQD molecule with staggered band alignment of the two cores. The electron and hole wavefunctions being in two different dots reduces the overlap between them, would make the charge carriers longer lived.

(d) The systematic change in the barrier, simply by changing the shell thickness i.e. yardstick for the center-to-center separation, brands the fused CQD molecules as ideal candidate to probe the coupling mechanism and underlying photophysics. At larger separation of the cores resonance energy transfer dominate the photophysical process while at shorter distance charge tunneling prevails. The inherent electric-field and the external perturbation of the same relies greatly on the degree and mode of coupling, where the photo-generated exciton can tunnel and switch between dots with modulation of e-field fostering CQD molecules for the utilization as electric field sensors.

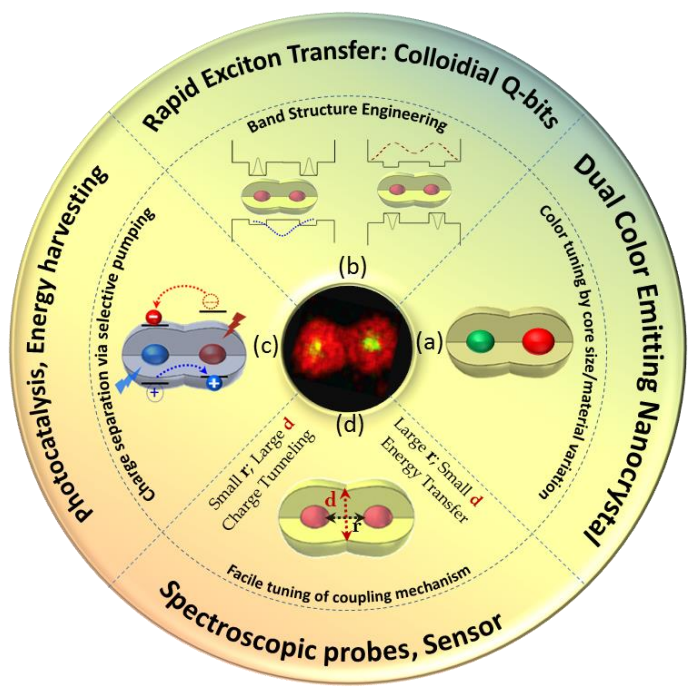

**Figure 7.** Library for rational design of parameters in fused core/shell dimer molecule towards potential applications. Systematic variation in the monomer units alter the extent and mode of coupling accordingly to achieve specific functionalities.

Apart from the directions and applications discussed, the hybridization of different atomic states such as the *s* and *p* states from different dots with comparable energy can also be hybridized which truly can construct the library of hybrid artificial molecules. In these systems, the coupling attributes in the coupled colloidal nanostructures including fused homodimer molecule and other particular heterostructures may be revealed at ambient conditions.

Feasible and targeted synthesis of colloidal nanocrystals with desired dimensionality and functionality has enriched nanoscience during the last decades. Complex architectures with hybrid optical and electrical properties manifest a contemporary and rapidly developing research field. In this accounts we have demonstrated how rational adjunction of nanocrystal artificial atoms offers the flexibility to control the hybridization of individual quantum states. All these rely on the ultimate level of understanding the complex interaction in the individual systems, manifesting a bright future for Colloidal Quantum Dot molecules.

**Author information**

**Affiliations**

Institute of Chemistry, The Hebrew University of Jerusalem, Jerusalem 91904, Israel.

The Center for Nanoscience and Nanotechnology, The Hebrew University of Jerusalem, Jerusalem 91904, Israel.

**Corresponding author**

Correspondence and requests for materials should be addressed to U.B. (uri.banin@mail.huji.ac.il)

**Author Contributions**

S.K., J.B.C. and Y.E.P. contributed equally to this work

**Acknowledgments**

The research leading to these results has received financial support from the European Research Council (ERC) under the European Union's Horizon 2020 research and innovation programme (grant agreement No [741767]). J.B.C and S.K acknowledge the support from the Planning and Budgeting Committee of the higher board of education in Israel through a fellowship. U.B. thanks the Alfred & Erica Larisch memorial chair. Y.E.P. acknowledges support by the Ministry of Science and Technology & the National Foundation for Applied and Engineering Sciences, Israel.